\documentclass[10pt]{iopart}
\usepackage{iopams, graphicx,url, bm, color, tikz, pgfplots, hyperref, times, xspace, soul, bm}
\usepackage{siunitx}
\expandafter\let\csname equation*\endcsname\relax
\expandafter\let\csname endequation*\endcsname\relax
\usepackage{amsmath, amsfonts, amssymb}
\usepackage{chemmacros}
\usepackage[switch]{lineno}
\usepackage[style=nature, backend=biber, sorting=none, url=true, doi=true, eprint=false, isbn=false, firstinits, maxcitenames=2, maxbibnames=4]{biblatex}

\let\originalleft\left
\let\originalright\right
\renewcommand{\left}{\mathopen{}\mathclose\bgroup\originalleft}
\renewcommand{\right}{\aftergroup\egroup\originalright}

\hypersetup{colorlinks=true, linkcolor=blue, citecolor=blue, urlcolor=blue}
\renewcommand\d{\ensuremath{\textrm{d}}}
\providecommand\kb{\ensuremath{k_{\text{B}}}\xspace}
\providecommand\ito{\^{I}to\xspace}
\providecommand\eavg{\ensuremath{E_{\text{avg}}}}
\providecommand\oi{\ch{O_{I}}\xspace}
\providecommand\oii{\ch{O_{II}}\xspace}
\providecommand\cii{\ch{C_{II}}\xspace}
\providecommand\e{\ch{e}\xspace}
\providecommand\otwo{\ch{O2}\xspace}
\providecommand\ntwo{\ch{N2}\xspace}

\renewcommand\ominus{\ch{O^-}\xspace}
\providecommand\co{\ch{CO}\xspace}
\providecommand\cotwo{\ch{CO2}\xspace}
\providecommand\cotwoplus{\ch{CO2^+}\xspace}
\providecommand\coplus{\ch{CO^+}\xspace}
\providecommand\sfsix{\ch{SF6}\xspace}
\providecommand\nitrile{\ch{C4F7N}\xspace}
\providecommand\ketone{\ch{C5F10O}\xspace}

\DeclareSourcemap{
  \maps[datatype=bibtex]{
    \map{
      \step[fieldset=abstract, null]
    }
  }
}

\DeclareSIUnit{\td}{Td}
\DeclareSIUnit{\torr}{Torr}

\begin{document}

\title{A 3D kinetic Monte Carlo study of streamer discharges in \ch{CO2}}

\author{R Marskar}
\address{SINTEF Energy Research, Sem S\ae lands vei 11, 7034 Trondheim, Norway.}
\ead{robert.marskar@sintef.no}

\begin{abstract}
  We theoretically study the inception and propagation of positive and negative streamers in \cotwo.
  Our study is done in 3D, using a newly formulated kinetic Monte Carlo discharge model where the electrons are described as drifting and diffusing particles that adhere to the local field approximation.
  Our emphasis lies on electron attachment and photoionization.
  For negative streamers we find that dissociative attachment in the streamer channels leads to appearance of localized segments of increased electric fields, while an analogous feature is not observed for positive-polarity discharges.
  Positive streamers, unlike negative streamers, require free electrons ahead of them in order to propagate.
  In \cotwo, just as in air, these electrons are supplied through photoionization.
  However, ionizing radiation in \cotwo is absorbed quite rapidly and is also weaker than in air, which has important ramifications for the emerging positive streamer morphology (radius, velocity, and fields).
  We perform a computational analysis which shows that positive streamers can propagate due to photoionization in \cotwo.
  Conversely, photoionization has no affect on negative streamer fronts, but plays a major role in the coupling between negative streamers and the cathode.
  Photoionization in \cotwo is therefore important for the propagation of both positive and negative streamers.
  Our results are relevant in several applications, e.g., \cotwo conversion and high-voltage technology (where \cotwo is used in pure form or admixed with other gases).
\end{abstract}

\submitto{\PSST}

\maketitle

\ioptwocol

\section{Introduction}
As with other gases, electric discharges in \cotwo begin with one or more initial electrons that accelerate in an electric field.
If the electron velocity becomes sufficiently high, collisions with \cotwo molecules lead to net ionization when the ionization probability exceeds the attachment probability.
As the process cascades through further ionization by acceleration of secondary electrons, build-up of space charge from the electrons and residual ions modifies the electric field in which the electrons originally accelerated.
This modification marks the onset of a streamer discharge \cite{Nijdam2020}, which is a filamentary type of low-temperature plasma.
Streamers have the peculiar property that they continuously modify the electric field in which they propagate, and thus exhibit a substantial degree of self-propagation.

Streamer discharges are categorized as positive or negative, depending on their direction of propagation relative to the electric field.
Negative streamers propagate in the direction of the electrons (hence opposite to the electric field), and are characterized by a negative space charge layer surrounding their channels.
Streamers that propagate opposite to the electron drift direction are called positive streamers, and unlike negative streamers they require a source of free electrons ahead of them.
In air and other \ntwo-\otwo mixtures, this source is photoionization.
\cotwo is another molecule that is relevant in multiple fields of research involving electrical discharges.
In high-voltage (HV) technology, for example, manufacturers of HV equipment are currently transitioning from the usage of \sfsix to environmentally friendlier alternatives, such as pure \cotwo or mixtures of \cotwo and \nitrile (also relevant are mixtures of air and \ketone).
However, photoionization in \cotwo is known to be much weaker than in air (for an overview, see \textcite{Pancheshnyi2015}).
It is now also accepted that photoionization sensitively affects the morphology of positive streamers in air \cite{Bagheri_2019, Marskar2020, Wang2023} since it produces electron-ion pairs in regions where the plasma density is low, which exacerbates noise at the streamer front.
Positive streamer branching thus occurs much more frequently in gases with lower amounts of photoionization \cite{Briels2008}.
Since photoionization in \cotwo is lower than in air, one may expect that positive streamer discharges propagate quite irregularly.

Few experimental studies have addressed streamer propagation in \cotwo.
Experiments by \textcite{Seeger2017} showed that the DC breakdown voltage of \cotwo is different for positive and negative polarities.
The authors investigated DC discharges in non-uniform fields for both polarities, and showed that the breakdown voltage for positive polarity is lower than for negative polarity at pressures $p\leq \SI{1}{\bar}$.
At higher pressures $p> \SI{1}{\bar}$ this trend was reversed, and breakdown at negative polarity consistently occured at a lower applied voltage than breakdowns at positive polarity.
This behavior is quite unlike that of air, where positive streamers propagate more easily than negative streamers over a wide range of pressures.
Large statistical time lags were also observed for positive streamers, but not for negative streamers.
Inception did not always occur for positive streamers, despite waiting times up to several minutes, indicating that initiatory electrons are quite rare in \cotwo.
No similar effect was reported for negative streamers, which suggests that the source of the initiatory electron could be different for the two polarities.
A more thorough investigation of inception times in \cotwo was recently presented by \textcite{Mirpour2022}, who investigated pulsed discharges with \SI{10}{\hertz} repetition rates. 

Theoretically, \textcite{Levko2017} studied streamer propagation in \cotwo gas using a Particle-In-Cell (PIC) model with Monte Carlo Collisions (MCC), ignoring photoionization and elucidating the intricate details of the electron velocity distribution.
\textcite{Bagheri2020} studied positive streamers in \cotwo and air using a fluid model.
For \cotwo, the authors claim that photoionization is an irrelevant mechanism, and in the computer simulations they replace it by a uniform background ionization.
The above theoretical studies were done in Cartesian 2D \cite{Levko2017} and axisymmetric 2D \cite{Bagheri2020}, and 3D simulations have not yet been reported.

In this paper we study the formation of positive and negative streamer discharges in pure \cotwo, in full 3D.
Our focus lies on the emerging morphology of the streamers, and in particular on the roles of electron attachment and photoionization.
We show that currently reported photoionization levels for \cotwo \cite{Pancheshnyi2015} can facilitate positive streamer propagation.
As we artificially decrease the level of photoionization, we find that higher voltages are required in order to initiate positive streamer discharges.
Negative streamers are also examined, and we show that the comparatively low levels of photoionization in \cotwo has virtually no effect on the dynamics of negative streamer heads.
However, photoionization is shown to play a role in the coupling of the negative streamer to the cathode.

This paper is organized as follows:
Our computational model is presented in \sref{sec:Model}, where we include the physical model and a brief overview of the numerical discretization that we use.
Results are presented in \sref{sec:results_negative} and \sref{sec:results_negative_photoi} for negative streamers, and in \sref{sec:results_positive} and \sref{sec:results_positive_photoi} for positive streamers.
The paper is then concluded in \sref{sec:conclusion}.

\section{Theoretical model}
\label{sec:Model}

\subsection{Physical model}
We use a physical model where the electrons are described as microscopic particles that drift and diffuse according to the local field approximation (LFA), i.e., we use a microscopic drift-diffusion model rather than a fluid drift-diffusion model \cite{Marskar2023Preprint}.
The transport equation for the electrons occurs in the form of \ito diffusion:

\begin{equation}
  \label{eq:species}
  \d\bm{X} = \bm{V}\d t + \sqrt{2D\d t}\bm{N},
\end{equation}
where $\bm{X}(t)$ is the electron position, and $\bm{V}$ and $D$ are the electron drift velocity and diffusion coefficient.
$\bm{N}$ indicates a normal distribution with a mean value of zero and a standard deviation of one, and we close the velocity relation in the LFA as

\begin{eqnarray}
  \bm{V} &= \bm{v}_\e\left(\bm{X}\right), \\
  D &= D_\e\left(\bm{X}\right),
\end{eqnarray}
where $\bm{v}_\e = -\mu_\e\bm{E}$ is the fluid drift velocity where $\mu_\e$ is the electron mobility and $\bm{E}$ is the electric field, and $D_\e$ is the fluid diffusion coefficient.
Our model is quite similar to a conventional macroscopic drift-diffusion model, except that we replace the electron transport kernel by a microscopic drift-diffusion process (i.e., an \ito process) and the reactions by a kinetic Monte Carlo (KMC) algorithm.
Further details regarding the \ito-KMC algorithm and its association with fluid drift-diffusion models is given in \cite{Marskar2023Preprint}.

We use a fluid drift-diffusion model for ions, whose densities are indicated by $n_i$ where $i$ is some species index.
The equation of motion for the ions is

\begin{equation}
  \frac{\partial n_i}{\partial t} = -\nabla\cdot\left(\bm{v}_i n_i - D_i\nabla n_i\right) + S_i.
\end{equation}
where $\bm{v}_i$, $D_i$, and $S_i$ are the drift velocities, diffusion coefficients, and source terms for ions of type of $i$.
The electric field $\bm{E}=-\nabla\Phi$ is obtained by solving the Poisson equation for the potential $\Phi$:

\begin{equation}
  \label{eq:poisson}
  \nabla^2\Phi = -\frac{\rho}{\epsilon_0},
\end{equation}
where $\rho$ is the space charge density and $\epsilon_0$ is the vacuum permittivity.

\subsection{Chemistry}
We consider a comparatively simple kinetic scheme for \cotwo consisting only of ionizing, attaching, and photon-producing reactions, see \tref{tab:co2_reactions}.
Excited states of \cotwo are not tracked in our model as we are presently only interested in the main ion products.
We also remark that while the KMC algorithm uses chemical propensities rather than the more conventional reaction rate coefficient, all reactions in this paper are first-order reactions and in this case the reaction rates in the KMC and fluid formulations are numerically equivalent.
The connection between the rates that occur in the chemical propensities and conventional reaction rates can otherwise be quite subtle for higher-order reactions, see e.g. \cite{Marskar2023Preprint,Marskar2023,Gillespie2007} for further details.

\begin{table}[h!t!b!]
  \centering
  \caption{
    \label{tab:co2_reactions}    
    \cotwo plasma chemistry used in this paper.
  }
  \begin{indented}
  \item[]
    \centering
    \begin{tabular}{@{}lll}  
      \hline
      Reaction & Rate & Ref. \\
      \hline
      $\e + \cotwo \rightarrow \e + \e + \cotwoplus$  & $k_\alpha\left(E/N\right)$ & \cite{Hagelaar2005a} \\
      $\e + \cotwo \rightarrow \co + \ominus$  & $k_\eta\left(E/N\right)$  & \cite{Hagelaar2005a} \\
      $\e + \cotwo \rightarrow \e + \cotwo + \gamma$  & $k_\gamma\left(E/N\right)$  & \cite{Hagelaar2005a} \\            
      \hline      
    \end{tabular}
  \end{indented}    
\end{table}

Transport coefficients and reaction rates for the electrons are computed using BOLSIG+ \cite{Hagelaar2005a} and the Phelps database (retrieved Oct. 16, 2023) \cite{PhelpsDatabase}.
The ion mobility is set to \SI{2E-4}{\square\meter\per\volt\per\second}.
All rates are calculated at standard atmosphere, i.e. $N\approx = \SI[per-mode=reciprocal]{2.446E25}{\per\cubic\meter}$.

\subsection{Electron attachment in \cotwo}
\begin{figure}[htb]
  \centering
  \includegraphics{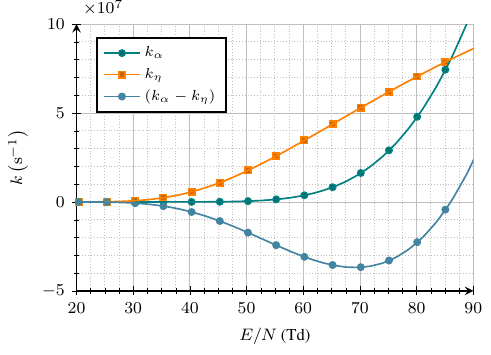}
  \caption{
    Ionization and attachment rate coefficients as functions of $E/N$ at atmospheric pressure.
  }
  \label{fig:AttachmentInstability}
\end{figure}

In the transport data we notice a peculiar feature that is relevant on longer timescales (tens of nanoseconds).
\Fref{fig:AttachmentInstability} shows the ionization and attachment coefficients ($k_\alpha$ and $k_\eta$), and the effective ionization rate $\left(k_\alpha-k_\eta\right)$ for the attachment region $E/N \leq \SI{90}{\td}$.
For fields $E/N\approx\SI{70}{\td}$ there is a global minimum in the effective ionization coefficient where dissociative attachment is particularly effective.
At atmospheric pressure, which is what we study, the attachment lifetime at $E/N\approx\SI{70}{\td}$ is $1/\left|k_\alpha -k_\eta\right|\approx \SI{20}{\nano\second}$.
If such fields appear in the streamer channel, dissociative attachment can potentially reduce the electron density by a factor of $1/\text{e}$ every \SI{20}{\nano\second}.
We mention this feature because an analogous phenomenon exists for streamer discharges in air, where it is known as the attachment instability \cite{DouglasHamilton1973}.

The physical explanation of the attachment instability is based on the tendency of streamer channels to become quasi-stationary due to the short relaxation time of the channel, in which case the current through the channel is constant \cite{Luque2016}.
We then obtain $\nabla\cdot\left(\sigma \bm{E}\right) = 0$ for the current density $\bm{J} = \sigma\bm{E}$, where $\sigma$ is the electric conductivity. 
When electron attachment reduces the conductivity the channel responds by increasing the electric field such that the current through the channel remains constant.
Because the effective attachment rate is field dependent with a local maximum around $E/N\approx\SI{70}{\td}$, this process is self-reinforcing.
Suppose for a moment that some region in the streamer channel initially has an internal field $E/N=\SI{50}{\td}$.
When dissociative attachment sets in, the field in the channel will start to increase as the conductivity is reduced.
However, as we move rightwards in \fref{fig:AttachmentInstability} from $E/N=\SI{50}{\td}$ the effective attachment rate increases further, which simply accelerates the rate of dissociative attachment and thus increases the field in the channel further. 
In recent calculations we showed that this mechanism is responsible for column glows and beads in so-called sprite discharges in the Earth atmosphere \cite{marskar2023genesis}.
\textcite{Malagon2019} also propose that the attachment instability is the reason why pilot systems \cite{Kochkin2016b} and space leaders \cite{Kochkin2014} appear in metre-scale discharges, as they lead to optical emission and heating of localized segments of the streamer channel.
However, it is not yet clear under which conditions the attachment instability begins to manifests since it requires a comparatively high initial electric field, well above the fields commonly observed in unperturbed streamer channels (at least for positive streamers in air).
Our transport data nonetheless suggests that an attachment instability is also present in \cotwo, which is of particular relevance to laboratory discharges as well as sprites in the Venusian atmosphere (which is mostly composed of \cotwo).

\subsection{Photoionization in \cotwo}
\label{sec:photoionization}
In contrast to the case of air where photoionization data is abundant and the primary states involved in the emission process have been identified, photoionization data for \cotwo is scarce.
Photoionization in air primarily occurs due to a Penning effect where \ntwo is first excited to the Carrol-Yoshino and Birge-Hopfield II bands, which have excitation energies higher than the ionization potential of \otwo.
The de-excitation pathways from excited \ntwo are collisional relaxation (i.e., collisional quenching), and spontaneous emission.
In the context of air, spontaneous emission rates are found in \cite{Stephens2018} (predissociation is also a relevant relaxation mechanism for \ntwo).
When excited \ntwo emits radiation through spontaneous emission it can ionize \otwo, and this supplies an efficient photoionization mechanism that produces free electrons.
However, this mechanism relies on the availability of two molecular components with different ionization potentials, so there can be no pure effect like this in single-component gases like pure \cotwo.
Photoionization in pure molecular gases must accordingly proceed first by formation of excited dissociation products, which is then followed by spontaneous emission of ionizing radiation.
In \cotwo, this may occur due to emission from \oi, \oii, \cii, \co, and \coplus \cite{Kanik1993}.
Emission from these fragments, which form due to dissociative excitation of \cotwo, can thus ionize \cotwo which has an ionization potential corresponding to \SI{90.5}{\nano\meter} radiation.

The emission cross sections for the dissociative fragments (\oi, \oii, \cii, \co, \coplus) that produce extreme ultraviolet (EUV) ionizing radiation below \SI{90.5}{\nano\meter} are incomplete, which prevents us from using cross sections when deriving a photoionization model.
\textcite{Kanik1993} provide emission cross sections for \SI{200}{\electronvolt} electrons and identify spectral peaks corresponding to emissions from \oi, \oii, \cii, \co, and \coplus.
For the \SI{83.4}{\nano\meter} peak which corresponds to emission from \oii, the authors also present energy-resolved cross sections.
The data in \textcite{Kanik1993} is not available in tabulated form, but for an electron energy of \SI{76.5}{\electronvolt} we may extract an approximate emission cross section of $\SI{1.78E-20}{\square\centi\meter}$ (from figure 2 in \cite{Kanik1993}).
The corresponding ionization cross section that we use at \SI{76.5}{\electronvolt} energy is approximately \SI{3.5E-16}{\square\centi\meter}, so the production of \oii emissions is considerably lower than the rate of electron impact ionization.
Unfortunately, the precision in the figures by \textcite{Kanik1993} makes it difficult to extract cross sections at lower electron energies and, furthermore, energy resolved cross sections are not available for the other EUV emissions.

The only available experiments that provide data for photoionization in \cotwo are due to \textcite{Przybylski1962} who performed experiments at pressures of \SIrange{1}{3}{\torr}.
Collisional quenching is most likely negligible at these pressures, and we have not been able to obtain data that describes the quenching rates of the involved EUV-emitting fragments.
Even in air, quenching rates for the Carrol-Yoshino and Birge-Hopfield II bands of \ntwo are not known individually (collisional de-excitation may occur at different rates for the two bands), but one may describe quenching by an approximate quenching pressure $p_q \sim \SI{40}{\milli\bar}$.
This leads to a correction in the photoionization level by $p_q/(p + p_q) \sim 0.04$ at \SI{1}{\bar} gas pressure, and this approach describes experiments with an acceptable level of accuracy \cite{Wang2023}.

\begin{figure}[htb]
  \centering
  \includegraphics{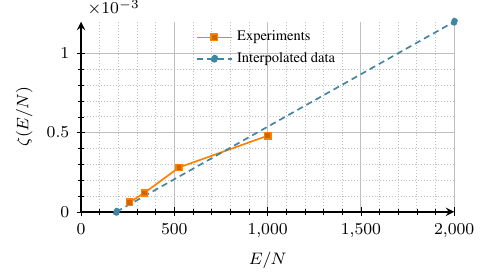}
  \caption{
    Coefficient $\xi(E/N)$, describing the number of ionizing photons produced per electron impact ionization event.
  }
  \label{fig:Photoyield}
\end{figure}

The present situation for \cotwo is not ideal: Appropriate energy-resolved emission cross sections at relevant electron energies are not available, and experimental data is only available for low-pressure \cotwo.
The atomic fragments that emit the EUV radiation might be quenched differently, implying that collisional quenching does not only reduce the number of ionizing photons, but potentially also their spatial distribution.
As we do not know of any data that provides an equivalent quenching pressure in \cotwo, we introduce a free parameter $\nu_q$ that adjusts the amount of photoionization in our simulations, which is to be interpreted as follows:
The quenching behavior of the EUV-emitting fragments (e.g., \oi) following impact dissociation of \cotwo obeys

\begin{equation}
  \partial_t \oi = -\frac{\oi}{\tau} - (k_q N)\oi,
\end{equation}
where $\tau$ is the radiative lifetime and $k_qN$ is the quenching rate.
Quenching occurs due to collisions between \oi and neutral \cotwo molecules, so the quenching rate grows linearly with neutral density $N$.
The number of photoemission events per de-excitation of \oi is then $\frac{\tau^{-1}}{\tau^{-1} + k_qN}$, and as $N$ is proportional to pressure ($p=N \kb T$), collisional quenching can reduce the amount of photoionization at higher pressures.
Similar relations could be formulated for the other fragments, but as none of the corresponding rate constants ($\tau^{-1}$ and $k_q$) are known, we lump this factor into a single term $\nu_q$.

The photon production rate $k_\gamma$ in our calculations is then calculated as

\begin{equation}
  k_\gamma(\nu_q) = \nu_q \xi\left(E/N\right) k_\alpha,
\end{equation}
where $\nu_q  \leq 1$ phenomenologically describes a reduction in the production of ionizing photons due to collisional quenching, and $\xi(E/N)$ is a field-dependent proportionality factor that describes the number of photoionization events per electron impact ionization event as originally measured by \textcite{Przybylski1962}.
For air, $\xi\left(E/N\right)$ is approximately $0.06$, while for \cotwo the reported value is at least one order of magnitude smaller.
We have presented this data in \fref{fig:Photoyield} versus $E/N$.
The experimental data is limited to $E/N \in \left[\SI{220}{\td}, \SI{1000}{\td}\right]$, so we linearly extrapolate the data as indicated in the figure.
This extrapolation is done because we observe that very high fields develop in computer simulations with low values of $\nu_q$, while emission cross sections generally peak at around \SI{200}{\electronvolt} \cite{Kanik1993}.

\begin{figure}[htb]
  \centering
  \includegraphics{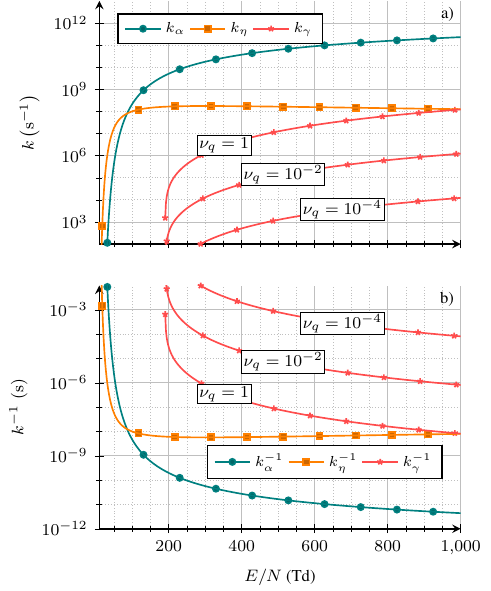}
  \caption{a) Ionization, attachment, and photon production rates for various quenching efficiencies $\nu_q$. b) Inverse rates (i.e., lifetimes).}
  \label{fig:Rates}
\end{figure}

\Fref{fig:Rates}a) shows the rates $k_\alpha$, $k_\eta$, and $k_\gamma(\nu_q)$ for $\nu_q=\numlist{E-4;E-2;1}$ as functions of the reduced electric field $E/N$.
We also include the inverse rates (i.e., lifetimes) of these reactions in \fref{fig:Rates}b).
The reaction lifetimes describe the average time before an electron triggers the reaction, and we can see that each electron generates one impact ionization collision every $k_\alpha^{-1}\sim \SI{10}{\pico\second}$ at $E/N\sim \SI{600}{\td}$.
However, the lifetime $k_\gamma\left(\nu_q=1\right)$ is approximately \SI{10}{\nano\second} at the same field strength, and photoionization events are thus rare compared to ionization events.

\cotwo absorbs quite strongly in the \SIrange{83}{89}{\nano\meter} spectral range, where the pressure-reduced mean absorption coefficient is between $\kappa_{\text{min}}/p = \SI{0.34}{\per\centi\meter\per\torr}$ and $\kappa_{\text{max}}/p = \SI{2.2}{\per\centi\meter\per\torr}$ \cite{Pancheshnyi2015}.
At atmospheric pressure this corresponds to mean photon absorption lengths between \SI{6}{\micro\meter} and \SI{38}{\micro\meter}.
This is shorter than in air, where mean absorption lengths are between \SI{30}{\micro\meter} and \SI{2}{\milli\meter} at atmospheric pressure.

When computational photons are generated in our simulations, their mean absorption coefficient is computed as

\begin{equation}
  \kappa = \kappa_{\text{min}}\left(\frac{\kappa_{\text{max}}}{\kappa_{\text{min}}}\right)^U,
\end{equation}
where $U$ is a random number sampled from a uniform distribution on the interval $[0,1]$, and $\kappa_{\text{min}}$ and $\kappa_{\text{max}}$ are as given above.
Only a few photons are generated per time step and cell.
A rough estimate may be obtained from \fref{fig:Rates}a) with $E/N=\SI{600}{\td}$, where $k_\gamma\sim\SI[per-mode=reciprocal]{2.75E5}{\per\second}$, while typical plasma densities $n_\e$ at streamer tips are \SIrange[per-mode=reciprocal]{E18}{E20}{\per\cubic\meter}.
Time steps are typically $\Delta t \sim k_\alpha^{-1} \approx \SI{10}{\pico\second}$ and grid cell volumes in the streamer head are $\Delta V \sim \SI{8E-18}{\cubic\meter}$.
The mean number of photons generated per cell and time step is roughly $k_\gamma n_\e \Delta t \Delta V$ which evaluates to between \num{2E-5} and \num{2E-3} photons on average.
Note that this estimate is per grid cell; the total number of ionizing photons emitted from a streamer head will be substantially higher.
We point out that the computational photons in our calculations correspond to physical photons, so there is no artificial elevation of discrete particle noise due to photoionization.

\subsection{Simulation conditions}
\begin{figure}[htb]
  \centering
  \includegraphics{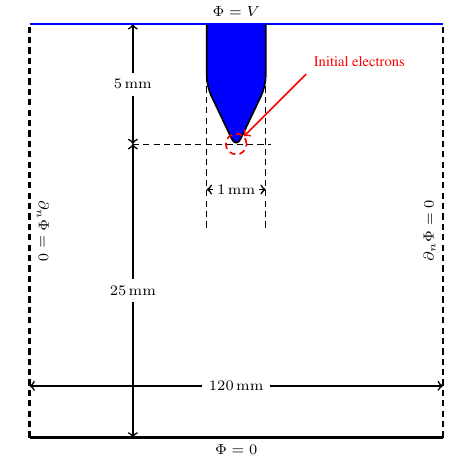}
  \caption{
    Sketch (not to scale) of the computational domain with electrostatic boundary conditions.
    The rounding radius at the needle tip is \SI{200}{\micro\meter} and the full opening angle of the conical section on the electrode is \num{30} degrees.
  }
  \label{fig:Domain}
\end{figure}

Our simulations are performed in the protrusion-plane geometry shown in \fref{fig:Domain}, which has dimensions of $\SI{12}{\centi\meter}\times\SI{3}{\centi\meter}\times\SI{12}{\centi\meter}$.
The discharges initiate at the tip of a \SI{5}{\milli\meter} long electrode that protrudes downwards along the $z$-axis.
The gap distance between the electrode and the ground plane is \SI{25}{\milli\meter}.
The protrusion radius is \SI{1}{\milli\meter}, and narrows along a conical section with a full opening angle of \num{30} degrees and a tip radius of \SI{200}{\micro\meter}.
All calculations are performed for a standard atmosphere (i.e., pressure $p=\SI{101325}{\pascal}$ and temperature $T=\SI{300}{\kelvin}$).

For the electric potential $\Phi$ we use homogeneous Neumann boundary conditions ($\partial_n\Phi = 0$) on the side faces and Dirichlet boundary conditions on the top and bottom faces.
The lower face is always grounded ($\Phi=0$) while the upper face and the protrusion are always at live voltage $\Phi=V$ where $V$ is a constant applied voltage which is varied in our computer simulations.
We also define the average electric field between the electrode and the ground plane as

\begin{equation}
  \eavg = \frac{V}{L},
\end{equation}
where $L=\SI{25}{\milli\meter}$.
The baseline quenching efficiency that we use in our simulations is $\nu_q=\num{1}$, but we vary this in \sref{sec:results_negative_photoi} and \sref{sec:results_positive_photoi}.

All simulations begin by sampling \num{100} physical electrons with random positions inside a \SI{200}{\micro\meter} radius sphere centered at the electrode tip.
Initial electrons whose positions end up inside the electrode are discarded before the simulation begins.
The initializing particles are unique to each simulation.

\subsection{Numerical discretization}
We use the chombo-discharge code \cite{Marskar2023} for performing our computer simulations.
As the full discretization and implementation of the model are quite elaborate, we only discuss the basic features here.

In time, we use a Godunov operator splitting between the plasma transport and reaction steps, where the transport step is semi-implicitly coupled to the electric field (see \cite{Ventzek1994} for another type semi-implicit coupling for fluid discretizations).
After the transport step we resolve the reactions in each grid cell using a KMC algorithm.
Unlike the deterministic reaction rate equation, the KMC algorithm is fully stochastic and operates with the number of particles in each grid cell rather than the particle densities.
Complete details are given in \cite{Marskar2023Preprint}.
Constant time steps $\Delta t = \SI{10}{\pico\second}$ are used in our simulations.

In space, we use an adaptive Cartesian grid with an embedded boundary (EB) formalism for solid boundaries.
EB discretization, also known as cut-cell discretization, is a special form of boundary discretization and brings substantial complexity into the discretization schemes (e.g., see \cite{Marskar2019b}).
In return it permits use of almost regular Cartesian data structures, and allows us to apply adaptive mesh refinement (AMR) in the presence of complex geometries with comparatively low numerical overhead.
Special handling of discretization routines is introduced at cut-cell and refinement boundaries.
For example, we always enforce flux matching for the Poisson equation \cite{Marskar2019}, and particles are deposited using custom deposition methods near refinement boundaries \cite{Marskar2023Preprint}.

We discretize the 3D domain using $\num{256}\times\num{64}\times\num{256}$ cells and add 9 levels of grid refinement which are dynamically adapted as simulations proceed.
The refinement factor between adjacent grid levels is always 2, so the finest representable grid cell in our simulation is $\Delta x \approx \SI{0.91}{\micro\meter}$.
Grid cells are refined every 5 time steps (i.e., every \SI{50}{\pico\second}) if

\begin{equation}
  \overline{\alpha} \Delta x \geq 1,
\end{equation}
where $\overline{\alpha}$ is the effective Townsend ionization coefficient.
Likewise, grid cells are coarsened if

\begin{equation}
  \overline{\alpha} \Delta x \leq 0.1 
\end{equation}
and $\Delta x$ was no larger than \SI{8}{\micro\meter}.

\begin{figure*}[h!t!b!]
  \centering
  \includegraphics{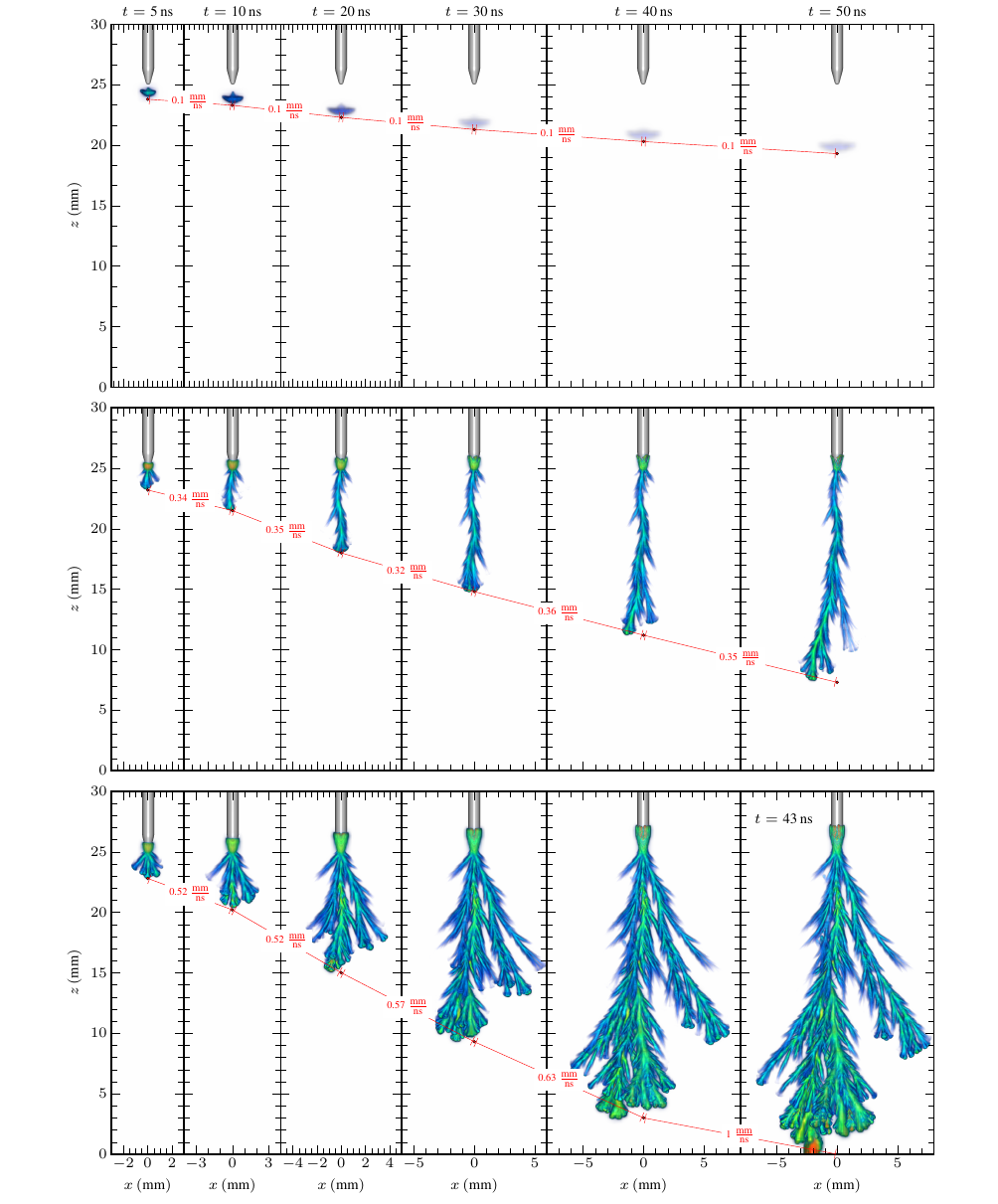}  
  \caption{Negative streamer evolution using various applied voltages.
    Top row: $V=\SI{-25}{\kilo\volt}$ ($\eavg =  \SI{1}{\kilo\volt\per\milli\meter}$).
    Middle row: $V=\SI{-30}{\kilo\volt}$ ($\eavg =  \SI{1.2}{\kilo\volt\per\milli\meter}$).
    Bottom row: $V=\SI{-35}{\kilo\volt}$ ($\eavg =  \SI{1.4}{\kilo\volt\per\milli\meter}$), where the final frame shows a snapshot when the streamer connects to the ground plane after $t=\SI{43}{\nano\second}$.
    The dashed lines shows the average vertical velocities (between the indicated markers in each figure).
  }
  \label{fig:NegativeVersusVoltage}
\end{figure*}

The \ito-KMC model we use is a particle-based model for the electrons, and particle re-balancing is required since the number of physical electrons at the streamer tips grows exponentially in time.
Particle merging and splitting is done following our previous approach discussed in \cite{Marskar2023Preprint} where bounding volume hierarchies are used for group partitioning of particles within a grid cell.
The algorithm is run at every time step, and ensures that computational particle weights within a grid cell differ by at most one physical particle.
In all simulations we limited the maximum number of computational particles to \num{32}.

The calculations in this paper were performed on \numrange{8}{80} nodes on the Betzy supercomputer, where each node consists of dual AMD EPYC 7742 CPUs.
Each node has $\num{2}\times\num{64}$ CPU cores, corresponding to a total of \numrange{1024}{10240} CPU cores for the various simulations.
Meshes ranged up to \num{2.5E9} grid cells and \num{E10} computational particles, with various simulations completing in \numrange{0.5}{5} days.

\section{Results}

\subsection{Negative streamers versus voltage}
\label{sec:results_negative}
In this section we present results for the evolution of negative streamers for voltages $V\in\left[-\SI{25}{\kilo\volt}, -\SI{30}{\kilo\volt}, -\SI{35}{\kilo\volt}\right]$.
These voltages corresponds to average electric fields of \SI{1}{\kilo\volt\per\milli\meter}, \SI{1.2}{\kilo\volt\per\milli\meter}, and \SI{1.4}{\kilo\volt\per\milli\meter}.
We performed a single 3D simulation for each voltage, and present the results in \fref{fig:NegativeVersusVoltage}.

\subsubsection{Streamer propagation field}
The top row in \fref{fig:NegativeVersusVoltage} shows that a negative streamer started at a voltage of $V=\SI{-25}{\kilo\volt}$, but the discharge did not propagate very far during the \SI{50}{\nano\second} simulation time.
The streamer was not electrically connected (with plasma) to the cathode either, and propagated as a diffuse cloud of electrons that gradually broadened and weakened with propagation distance.
While we did not run the simulation further, we expect that the discharge would eventually fade out and decay.
The middle row in \fref{fig:NegativeVersusVoltage} shows negative streamers at a voltage of $V=\SI{-30}{\kilo\volt}$, corresponding to an average electric field of $\eavg=\SI{1.2}{\kilo\volt\per\milli\meter}$.
The streamer is characterized by a main branch with numerous side branches, many of which stagnate early and do not propagate further. 
This streamer did not cross the discharge gap in the course of the simulation, although we expect that it would have if the simulation was run further.
Finally, the bottom row shows the streamer development with an applied voltage of $V=\SI{-35}{\kilo\volt\per\milli\meter}$, corresponding to $\eavg=\SI{1.4}{\kilo\volt\per\milli\meter}$.
The discharge consists of multiple branches that form a broad discharge tree approximately \SI{15}{\milli\meter} wide, and crossed the discharge gap in \SI{43}{\nano\second}.

From the results we conclude that negative streamers in our simulations propagate if the average electric field is $\eavg \in \SIrange{1}{1.2}{\kilo\volt\per\milli\meter}$.
Incidentally, \textcite{Seeger2017} report that the negative streamer stability field derived from their experiments is \SI[separate-uncertainty=true]{11\pm 2}{\volt\per\meter\per\pascal}, which translates to \SI[separate-uncertainty=true]{11\pm2}{\kilo\volt\per\milli\meter} at \SI{1}{\bar} gas pressure.
We thus find quantitative agreement between experiments \cite{Seeger2017} and our observed propagation fields.

\subsubsection{Velocity}
In the computer simulations we observe that the front velocity of the discharge is voltage dependent, and that it varies during the discharge evolution.
We have indicated the velocities in \fref{fig:NegativeVersusVoltage}, which are calculated by estimating how far the vertical front position of the discharge has moved between the frames.
At the lowest voltage ($V=\SI{-25}{\kilo\volt}$) the discharge propagated with an average velocity of \SI{0.1}{\milli\meter\per\nano\second}, but as we mentioned above the discharge does not represent a propagating streamer.
For $V=\SI{-30}{\kilo\volt}$ the observed velocity remained fairly constant throughout the propagation phase, with an approximate value of $v = \SIrange{0.34}{0.36}{\milli\meter\per\nano\second}$.
The bottom row in \fref{fig:NegativeVersusVoltage} shows that at the highest simulated voltage $V=\SI{-35}{\kilo\volt}$ the front velocity varied a great deal throughout the streamer development.
For $t<\SI{20}{\nano\second}$ the streamer velocity was approximately was $\SI{0.52}{\milli\meter\per\nano\second}$, which increased to approximately $\SI{1}{\milli\meter\per\nano\second}$ as the streamer approach the ground plane.

\textcite{Seeger2017} have measured approximate  streamer velocities in \cotwo, using a single PMT for estimating the propagation time of the streamer and an ICCD camera for measuring the streamer length.
The results for negative streamers were obtained for a field distribution slightly different from ours, and the authors report negative streamer velocities in the range of \SIrange{0.2}{0.6}{\milli\meter\per\nano\second} at \SI{1}{\bar} gas pressure.
We also point out that this velocity interval represents average streamer velocities rather than instantaneous velocities.
Our simulation results nonetheless agree quite well with the experimental values, despite the fact that the experimentally estimated streamer velocities contain uncertainties due to the measurement method.

\subsubsection{Radius}
\Fref{fig:NegativeVersusVoltage} shows that the negative streamers branch frequently, and many of the branches also stagnate, which makes it difficult to extract a single streamer radius in our calculations.
Since negative streamers can broaden quite efficiently, a range of streamer radii can probably be observed also in experiments.
In experiments, only the optical radius of the streamers are available, and \textcite{Seeger2017} report experimentally obtained negative streamer radii as \SI[separate-uncertainty=true]{25\pm3}{\meter\pascal}, which translates to \SI{250}{\micro\meter} at atmospheric pressure.
This radius was obtained for streamer filaments that did not branch, and thus correspond to the minimal streamer radius.
Our simulations do not model optical emission in the \cotwo plasma, so only the electrodynamic radius is available.
These measures can differ substantially.
For positive streamers in air it is estimated that the electrodynamic radius is twice that of the optical radius \cite{Pancheshnyi2005}, but no corresponding relation has been reported for negative streamers. 

\begin{figure}[htb]
  \centering
  \includegraphics{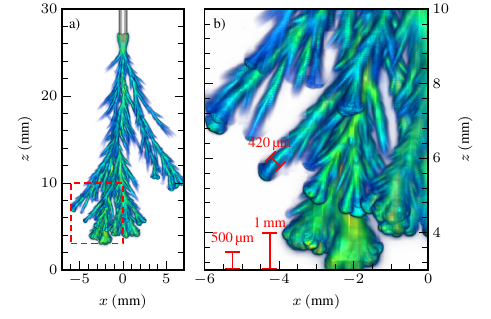}
  \caption{
    Determination of the minimum negative streamer radius for an applied voltage $V=\SI{-35}{\kilo\volt}$ after $t=\SI{40}{\nano\second}$, using the plasma density $n_\e$ as a proxy for the electrodynamic radius.
    a) Full view, showing plasma densities $n_\e\geq\SI[per-mode=reciprocal]{E18}{\per\cubic\meter}$.
    b) Inset of the indicated region in a).
  }
  \label{fig:NegativeRadius}
\end{figure}

\Fref{fig:NegativeRadius} shows the plasma density for the simulation with $V=\SI{-35}{\kilo\volt}$ after $t=\SI{40}{\nano\second}$.
In \fref{fig:NegativeRadius}b) we have also included various length indicators, as well as the diameter of a specific branch which initially propagated but later stagnated.
Examining the various branches in the figure we find that the smallest electrodynamic diameter of the filaments is at least \SI{420}{\micro\meter}, in good agreement with experimentally reported values \cite{Seeger2017}.

\begin{figure}[htb]
  \centering
  \includegraphics{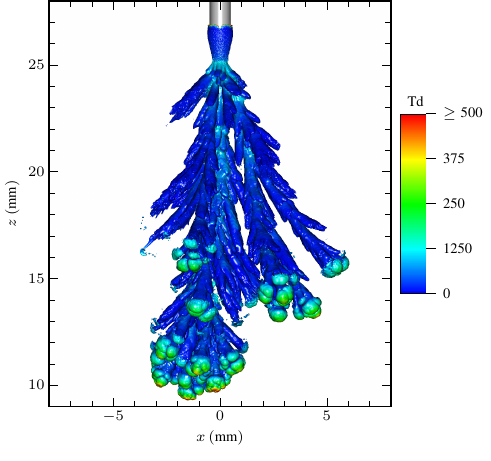}
  \caption{
    Isosurface $n_\e=\SI[per-mode=reciprocal]{E18}{\per\cubic\meter}$ for $V=-\SI{35}{\kilo\volt}$ after $t=\SI{30}{\nano\second}$.
    The surface is colored by the reduced electric field $E/N$.
  }
  \label{fig:NegativeField}
\end{figure}

\subsubsection{Field distribution}
In order to obtain an estimate for the range of electric fields that occur on negative streamer tips, \fref{fig:NegativeField} shows an isosurface $n_\e=\SI[per-mode=reciprocal]{E18}{\per\cubic\meter}$ of the plasma after $t=\SI{30}{\nano\second}$ for the simulation with $V=\SI{-35}{\kilo\volt}$.
Depending on the radius and position of the negative streamer tips, we find that the electric field at the negative streamer tips is \SIrange{250}{500}{\td}.
For comparison, reported electric fields for negative streamers in air are approximately \SI{300}{\td} \cite{Alejandro2008}.

\begin{figure}[htb]
  \centering
  \includegraphics{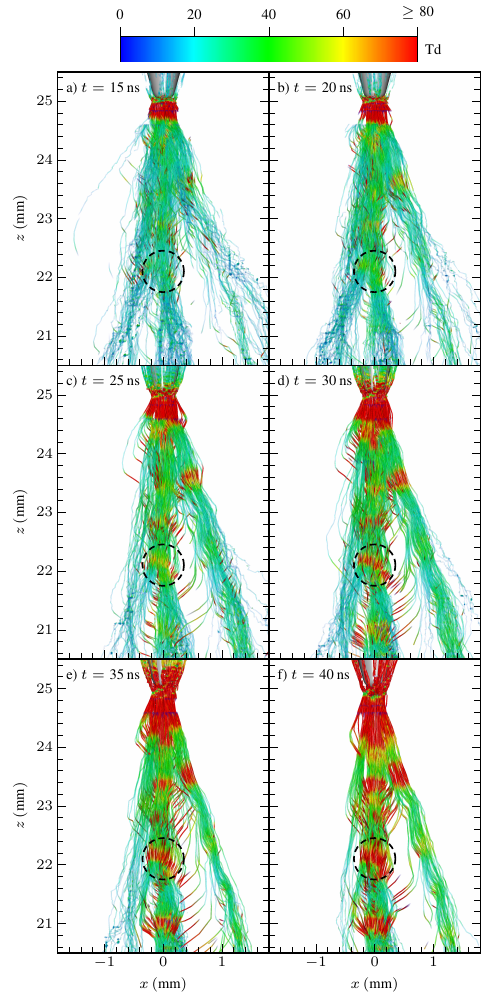}
  \caption{
    Field lines in the plasma colored by electric field (in units of \si{\td}, for $V=\SI{-35}{\kilo\volt}$ applied voltage.
    The color range is truncated to $E/N\in\SIrange{0}{80}{\td}$ with alpha channels that reduce the opacity of the field lines with lower $E/N$.
  }
  \label{fig:NegativeStreamlines}
\end{figure}

Next, we examine the evolution of the electric field in the streamer channels.
\Fref{fig:NegativeStreamlines} shows snapshots of electric field lines at various time instants for the simulation with $V=\SI{-35}{\kilo\volt}$.
Field lines are pruned from the plot if the electron density is $n_\e < \SI[per-mode=reciprocal]{E18}{\per\cubic\meter}$ so that all field lines in \fref{fig:NegativeStreamlines} pass through the plasma.
The field lines are colored by $E/N$, and transparency channels are added such that field lines with $E/N \geq \SI{80}{\td}$ are opaque and field lines with $E/N=\SI{0}{\td}$ are completely transparent.
\Fref{fig:NegativeStreamlines} shows that localized regions in the streamer channel with initially low electric fields later develop comparatively high fields $E/N > \SI{70}{\td}$.
We have indicated one of these regions by a dashed circle in \fref{fig:NegativeStreamlines}, but other regions can also be identified.
The field enhancement in the channels is caused by dissociative attachment which reduces the conductivity of the channel, as discussed in \sref{sec:Model}.
The conductivity reduction is then compensated by an increased electric field, similar to how the attachment instability operates in air \cite{DouglasHamilton1973}.

\begin{figure}[htb]
  \centering
  \includegraphics{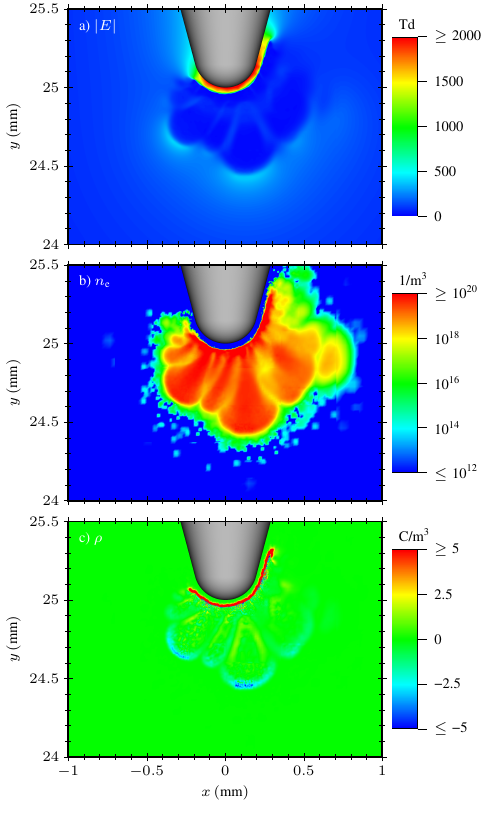}
  \caption{
    Cathode sheath region details at $t=\SI{2.5}{\nano\second}$.
    The data in each figure shows a slice through the $z$-plane in the simulation with $\nu_q=1$, $V=\SI{-35}{\kilo\volt}$.
    a) Electric field magnitude.
    b) Electron density.
    c) Space charge density.
  }
  \label{fig:CathodeSheath}
\end{figure}

\subsubsection{Cathode sheath}
\label{sec:CathodeSheath}

Negative streamers propagate away from the cathode and leave behind positive space charge composed of positive ions, which can lead to a sheath immediately outside of the cathode surface.
The sheath is electron-depleted because the electrons in it propagate away from the cathode (and thus out of the sheath).
Analogous sheaths also exist for positive streamers propagating over dielectric surfaces \cite{Meyer2020,Meyer2022,Marskar2023}.
Unfortunately, we can not study the details in the sheath with desired accuracy because of the inherent limitations of the LFA.
Physically, secondary electrons that appear in the sheath are due to photoionization, cathode emission, or electron impact ionization.
The secondary electrons arising from these processes are low-energy electrons that do not generate further impact ionization until they have been sufficiently accelerated to above-ionization energies.
But in LFA-based models these electrons are always born with artificially high energies, parametrically given as a function of $E/N$.
In our model, photoelectrons that appear in the sheath can thus immediately ionize the gas, which is non-physical since their true energy is $\mathcal{O}\left(\SI{1}{\electronvolt}\right)$.
Our model therefore predicts an artificially high level of impact ionization in the sheath region, and we can thus only make a qualitative assessment of the sheath features (such as its thickness).

\Fref{fig:CathodeSheath} shows some details of the cathode sheath region for the computer simulation with $\nu_q=1$, $V=\SI{-35}{\kilo\volt}$, where we include slice plots of the electric field magnitude, the electron density, and the space charge density.
From the figure we find a sheath thickness of approximately \SI{50}{\micro\meter}, and fields that range up to $\SI{2000}{\td}$.
\Fref{fig:CathodeSheath}c) shows the reason why this high field region appears, which is due to \cotwoplus ions that have accumulated just outside the cathode surface.
Since the cathode surface is charged negatively and the space charge is positive, there is a corresponding high field region between these two features.
This field will not persist indefinitely because as the ions move slowly towards the cathode, the space charge layer is gradually absorbed by the cathode and the field in the sheath will correspondingly decrease.
We have not shown this process in detail but point out that it occurs on a comparatively long time scale (tens to hundreds of nanoseconds) due to the comparatively low ion mobility.
For sheaths along dielectric surfaces, this can lead to charge saturation, as demonstrated by \textcite{8785920}.

\begin{figure}[htb]
  \centering
  \includegraphics{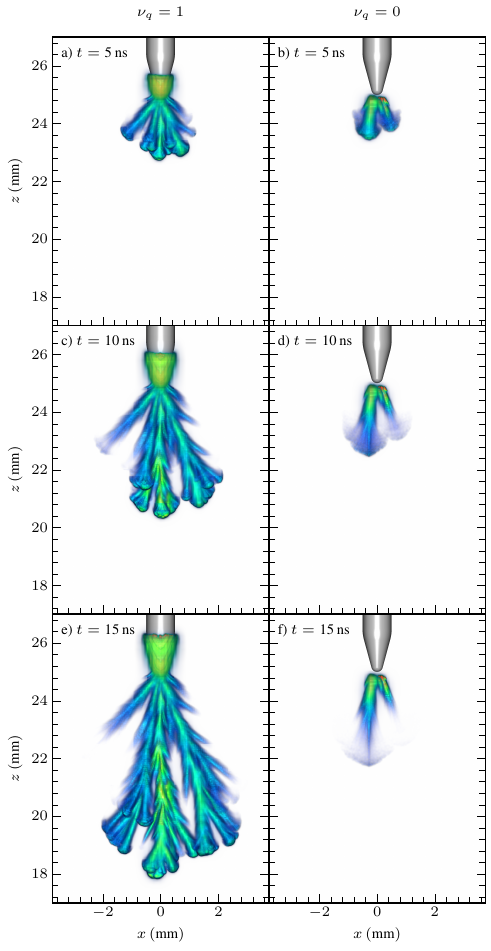}
  \caption{Evolution of a negative streamer with $V=\SI{-35}{\kilo\volt}$ with photoionization ($\nu_q=1$) and without photoionization ($\nu_q=0$).}
  \label{fig:NegativeNoPhotoi}  
\end{figure}

\subsection{Negative streamers without photoionization}
\label{sec:results_negative_photoi}
Few publications have addressed the role of photoionization in negative streamers.
Most of the available results are for air and with continuum approximations for the photons \cite{Luque2007,Starikovskiy2020}.
\textcite{Starikovskiy2020} provide a qualitative explanation on the role of photoionization for negative streamers:
Seed electrons that appear ahead of negative streamers turn into avalanches that propagate outwards from the streamer tip, which facilitates further expansion of the streamer head.
When the negative streamer head expands, field enhancement and thus impact ionization at the streamer tip decreases, which leads to a slower streamer.
Similar conclusions were reached by \textcite{Alejandro2008}, who also point out that this broadening can also lead to negative streamer decay (similar to \fref{fig:NegativeVersusVoltage} for $V=\SI{-25}{\kilo\volt}$).

It is important to note that the above cited results all use continuum approximations for photoionization, which is not a valid approximation for our conditions.
The role of discrete photoionization for positive streamers in air has been reported \cite{Bagheri_2019, Marskar2020}, and the studies show that positive streamer morphologies depend sensitively on the photoionization parameters.
Analogous studies for negative streamers in air have not yet been reported.
However, since photoionization can provide seed electrons ahead of negative streamers in precisely the same way as for positive streamers, photoionization might also play a role in the branching of negative streamers.

\Fref{fig:NegativeNoPhotoi} shows the plasma density for a case where photoionization is fully turned off ($\nu_q=0$).
The applied voltage is $V=\SI{-35}{\kilo\volt}$, i.e. corresponding to the bottom row in \fref{fig:NegativeVersusVoltage}, which is included for the sake of comparison.
Without photoionization, the negative streamer propagates much slower than its photoionization-enabled counterpart, and it eventually also decays.

The computer simulations also show that the cathode sheath dynamics are affected by photoionization.
\Fref{fig:NegativeNoPhotoi} shows that the cathode is partially covered by plasma when photoionization is enabled ($\nu_q=1$).
This plasma is a positive streamer that propagates upwards along the cathode, and it leaves behind a positive space charge layer outside the cathode (as seen in \fref{fig:CathodeSheath}).
The sheath is thus affected by the appearance of seed electrons in the cathode region, in particular seed electrons that appear in the cathode fall since these electrons initiate new avalanches that leave behind additional space charge.
As this process cascades, it leads to inception of a positive streamer that propagates towards and finally along the cathode surface.
Without photoionization, the necessary seed electrons that are required in order to faciliate the positive streamer no longer appear.
The upwards propagating positive streamer does not manifest without this source of electrons, and this reduces the intensity of the space charge layer close to the cathode and also the field in the cathode sheath.

\begin{figure}[htb]
  \centering
  \includegraphics{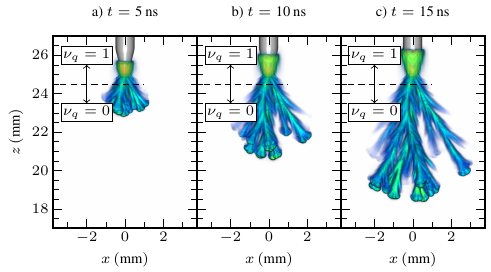}
  \caption{Evolution of a negative streamer with $V=\SI{-35}{\kilo\volt}$, using $\nu_q=1$ for $z>\SI{24.5}{\milli\meter}$ and $\nu_q=0$ elsewhere.}
  \label{fig:NegativePartialPhotoi}  
\end{figure}

\begin{figure*}[h!t!b!]
  \centering
  \includegraphics{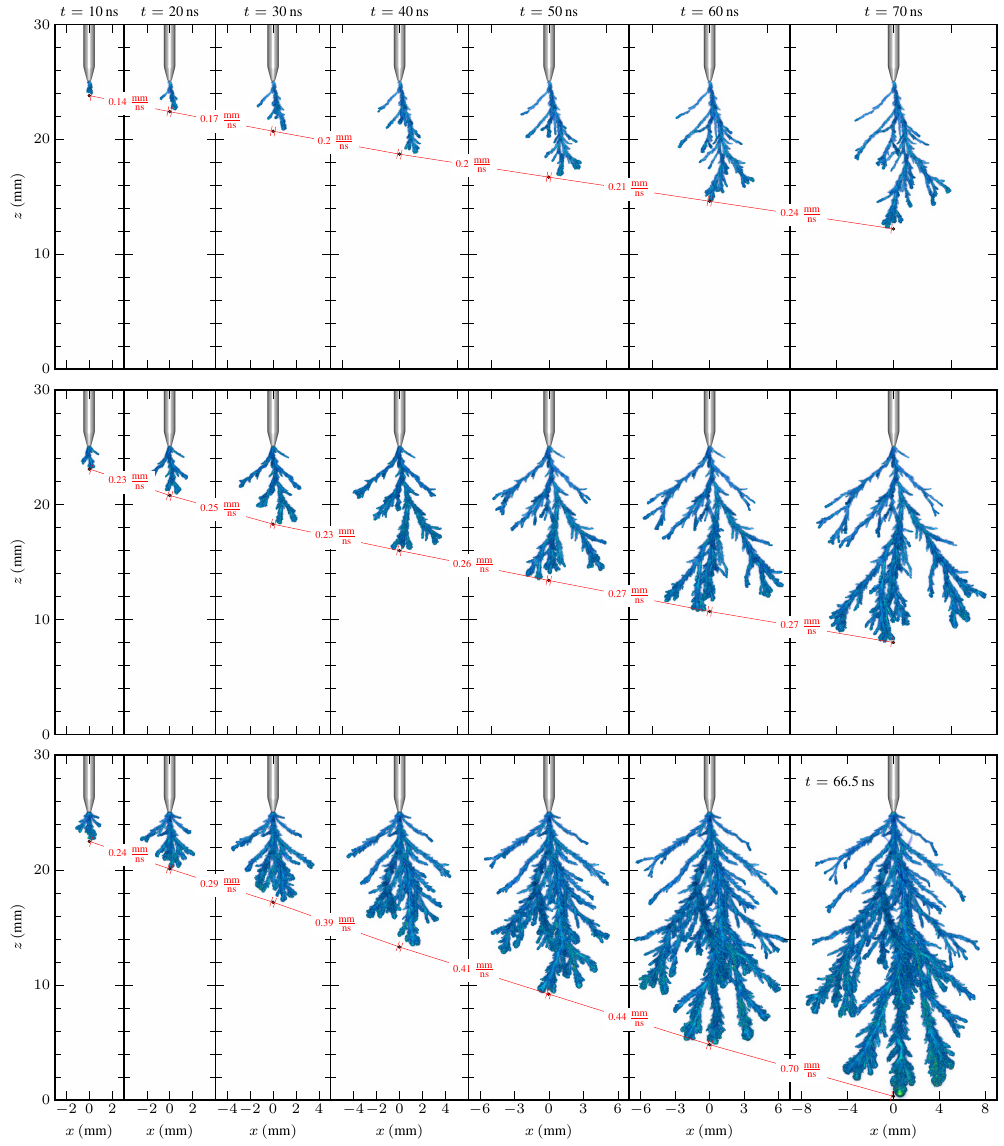}
  \caption{Positive streamer evolution with various applied voltages.
    Top row: $V=\SI{25}{\kilo\volt}$ ($\eavg = \SI{1}{\kilo\volt\per\milli\meter}$).
    Middle row: $V=\SI{30}{\kilo\volt}$ ($\eavg = \SI{1.2}{\kilo\volt\per\milli\meter}$).
    Bottom row: $V=\SI{35}{\kilo\volt}$ ($\eavg = \SI{1.4}{\kilo\volt\per\milli\meter}$).
    The final frame shows the plasma at $t=\SI{66.5}{\nano\second}$, immediately before the streamer connected to the ground plane.
    The dashed lines shows the average vertical velocities (between the indicated markers in each figure).
    }
  \label{fig:PositiveVersusVoltage}
\end{figure*}

While \fref{fig:NegativeNoPhotoi} shows that photoionization is important for the negative streamer evolution, it does not answer whether or not this is due to conditions at the negative streamer tip, or due to absence of the upwards positive streamer.
The positive streamer feeds a current into the system, and consequently it affects the potential distribution and field enhancement of the negative streamer head.

In order to determine whether or not the decay of the negative streamer seen in \fref{fig:NegativeNoPhotoi} is due to lack of photoionization at the negative streamer tip or absence of a positive cathode-directed streamer, we run another computer simulation where generation of ionizing photons is turned off in all regions where $z < \SI{24.5}{\milli\meter}$.
This is equivalent to turning off photoionization for the propagating negative streamer, but maintaining photoionization for the upwards positive streamer that propagates along the cathode surface.
\Fref{fig:NegativePartialPhotoi} shows the results for this simulation, and should be contrasted with the left and right columns in \fref{fig:NegativeNoPhotoi}.
The evolution of the negative streamer for this case is qualitatively similar to the left column of \fref{fig:NegativeNoPhotoi} where photoionization was enabled everywhere.
Consequently, photoionization at negative streamer tips in \cotwo does not appear to play a major role in the streamer evolution.
However, appropriate coupling to the cathode still requires inclusion of photoionization; the upwards positive streamer feeds a current into the tail of the negative streamer which increases field enhancement at the negative streamer tip, and thus facilitates its propagation.

\subsection{Positive streamers versus voltage}
\label{sec:results_positive}

In this section we consider propagation of positive streamers for voltages $V\in\left[\SI{20}{\kilo\volt}, \SI{25}{\kilo\volt}, \SI{30}{\kilo\volt}, \SI{35}{\kilo\volt}\right]$.
We perform the study in the same way as we did for negative streamers: A single computer simulation is performed for each voltage application, and we extract velocities, radii, and field distributions.
The evolution of the corresponding discharges is shown in \fref{fig:PositiveVersusVoltage}.

\subsubsection{Streamer propagation field}
The top row in \fref{fig:PositiveVersusVoltage} shows the evolution of positive streamers in \cotwo at $V=\SI{25}{\kilo\volt}$, which corresponds to an average electric field of $\eavg=\SI{1}{\kilo\volt\per\milli\meter}$.
We did not include the simulation with $V=\SI{20}{\kilo\volt}$ ($\eavg=\SI{0.8}{\kilo\volt\per\milli\meter}$) in \fref{fig:PositiveVersusVoltage} because a positive streamer failed to develop at this voltage.
The middle row in \fref{fig:PositiveVersusVoltage} shows the evolution for $V=\SI{30}{\kilo\volt}$ ($\eavg=\SI{1.2}{\kilo\volt\per\milli\meter}$) and the bottom row shows the evolution with $V=\SI{35}{\kilo\volt}$ ($\eavg = \SI{1.4}{\kilo\volt\per\milli\meter}$).
Qualitatively, we observe that with increasing voltage the streamers evolve into broader and faster discharge trees.
The discharges also grow much more irregularly than for negative polarity (see \fref{fig:NegativeVersusVoltage}).

Our baseline simulations show that positive streamers develop at $V=\SI{25}{\kilo\volt}$ but not $V=\SI{20}{\kilo\volt}$, indicating that the streamer propagation field in our calculations is \SIrange{0.8}{1}{\kilo\volt\per\milli\meter}.
This is, in fact, substantially lower than what is observed in experiments where the reported streamer propagation field at \SI{1}{\bar} is approximately \SI{1.3}{\kilo\volt\per\milli\meter} \cite{Seeger2017}.
For positive streamers, our calculations of the streamer propagation field therefore contain an error of about \SIrange{30}{50}{\percent}.

\subsubsection{Velocity}
Like we found with negative streamers, \fref{fig:PositiveVersusVoltage} shows that the front velocity of the discharge depends on the applied voltage and also that it varies during the discharge evolution. 
With $V= \SI{25}{\kilo\volt}$ the discharge propagated with an average velocity of approximately \SI{0.2}{\milli\meter\per\nano\second}, which is lower than the slowest negative streamer we observed ($V=\SI{-30}{\kilo\volt}$). 
For $V=\SI{30}{\kilo\volt}$ the observed velocity increased to approximately \SIrange{0.23}{0.27}{\milli\meter\per\nano\second}, while for $V=\SI{35}{\kilo\volt}$ the velocity was \SIrange{0.24}{0.44}{\milli\meter\per\nano\second}.
Our simulations show that positive streamers propagate slower than negative streamers, in agreement with experiments \cite{Seeger2017}.
For comparison, the corresponding average velocities deduced from experiments are \SIrange{0.1}{0.5}{\milli\meter\per\nano\second}, so our velocity calculations are in comparatively good agreement.

\subsubsection{Radius}
\begin{figure}[htb]
  \centering
  \includegraphics{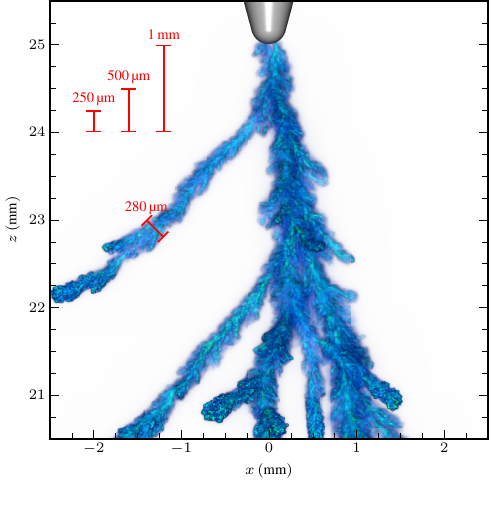}
  \caption{
    Determination of the minimum positive streamer radius for an applied voltage $V=\SI{25}{\kilo\volt}$ after $t=\SI{50}{\nano\second}$, using the plasma density $n_\e$ as a proxy for the electrodynamic radius.
  }
  \label{fig:PositiveRadius}
\end{figure}

\Fref{fig:PositiveVersusVoltage} shows that positive streamers in \cotwo can develop into tree structures that have a distribution of radii, i.e. there is no unique streamer radius.
As we did for negative streamers, we only extract the radius for streamer filaments that do not branch, using regions where $n_\e\geq \SI[per-mode=reciprocal]{E18}{\per\cubic\meter}$ as a proxy for the electrodynamic radius.
\Fref{fig:PositiveRadius} shows the simulation data for $V=\SI{25}{\kilo\volt}$ after $t=\SI{50}{\nano\second}$.
Various length indicators are included, as well as the diameter of a specific branch that did not branch, but whose path fluctuated.
From this branch we extract an approximate radius of $\SI{140}{\micro\meter}$.
This agrees quite well with the experiments by \textcite{Seeger2017} who report that the optical radius of positive streamers is at least \SI[separate-uncertainty=true]{13.4\pm3}{\meter\pascal}, which translates to \SI{130}{\micro\meter} at atmospheric pressure.

\subsubsection{Field distribution}
\begin{figure}[htb]
  \centering
  \includegraphics{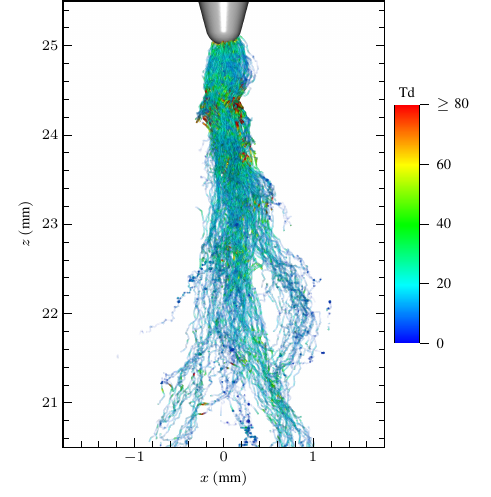}
  \caption{
    Field lines in the plasma colored by electric field (in units of \si{\td}, for $V=\SI{35}{\kilo\volt}$ applied voltage after $t=\SI{39}{\nano\second}$.
    The color range is truncated to $E/N\in\SIrange{0}{80}{\td}$ with alpha channels that reduce the opacity of the field lines with lower $E/N$.
  }
  \label{fig:PositiveStreamlines}
\end{figure}

As we discussed in \sref{sec:results_negative} we found that on longer timescales the field in negative streamer channels gradually increase due to an attachment instability that reduces the channel conductivity and hence increases the field in the channel.
We have not found any corresponding field increase in positive streamer channels, which suggests that the field is too low for the attachment instability to manifest at the positive streamer evolution time scale.
\Fref{fig:PositiveStreamlines} shows the field distribution along field lines in the streamer channels, i.e. regions where $n_\e\geq \SI[per-mode=reciprocal]{E18}{\per\cubic\meter}$.
In positive streamer channels we find that the internal electric field is \SIrange{10}{20}{\td}.
For comparison, this is the same value as positive streamers in atmospheric air, which is usually reported as being around \SI{20}{\td} \cite{Alejandro2008}.

\begin{figure}[htb]
  \centering
  \includegraphics{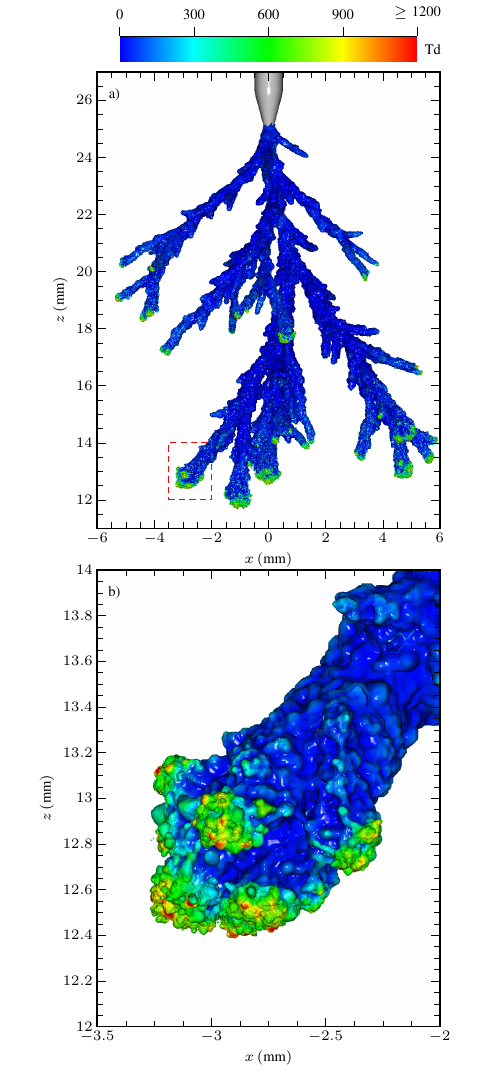}
  \caption{
    Isosurface $n_\e=\SI[per-mode=reciprocal]{E18}{\per\cubic\meter}$ for $V=\SI{30}{\kilo\volt}$ after $t=\SI{56}{\nano\second}$.
    The surface is colored by the reduced electric field $E/N$.    
    a) Full view.
    b) Closeup of the indicated region from a).
  }
  \label{fig:PositiveField}
\end{figure}

\Fref{fig:PositiveField} shows an isosurface $n_\e = \SI[per-mode=reciprocal]{E18}{\per\cubic\meter}$ for the simulation with $V=\SI{30}{\kilo\volt}$ after $t=\SI{56}{\nano\second}$.
The isosurface is colored by the reduced electric field $E/N$, and shows the reduced electric field at positive streamer tips.
Typical fields are \SIrange{600}{900}{\td}, but we point out that the positive streamer fronts are quite irregular with local enhancements of the field at their tips, as can be seen in \fref{fig:PositiveField}b) which shows a closeup near one of the positive streamer tips.
At this tip the field is locally enhanced to $E/N \geq \SI{1200}{\td}$, and the plasma irregularity at the tip can be identified.
We believe that this irregularity is caused by the low amount of photoionization, in which the incoming avalanches that grow toward the streamer tips lead to a fine-grained local field enhancement.

\begin{figure*}[htb]
  \centering
  \includegraphics{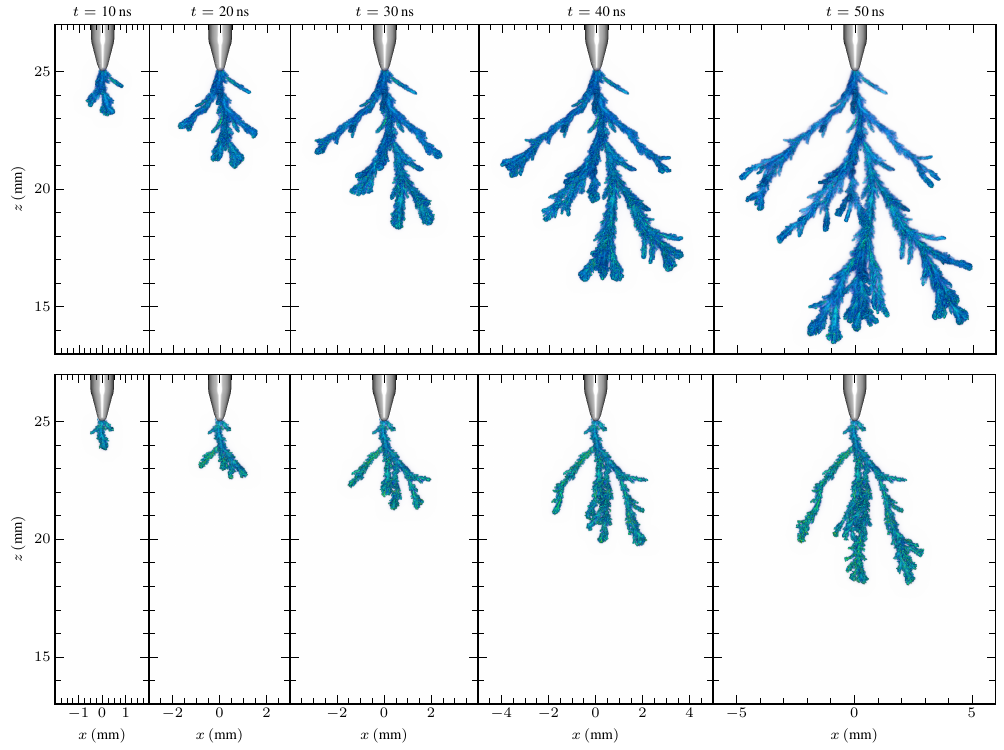}
  \caption{
    Comparison of positive streamers using an applied voltage $V=\SI{30}{\kilo\volt}$ with $\nu_q=\num{1}$ (top row) and $\nu_q=\num{0.1}$ (bottom row).
  }
  \label{fig:PositivePhotoi}
\end{figure*}

\subsection{Positive streamers with varying photoionization}
\label{sec:results_positive_photoi}
We now consider the evolution of positive streamers when we vary the amount of photoionization to $\nu_q=\num{0.1}$ and $\nu_q=\num{0.01}$.
We first ran simulations with $V=\SI{25}{\kilo\volt}$ using these parameters, but at this voltage the streamers failed to develop.
Recalling that the streamer propagation field was \SIrange{0.8}{1}{\kilo\volt\per\milli\meter} with $\nu_q=\num{1}$, we find that the streamer propagation field increases to at least \SIrange{1}{1.2}{\kilo\volt\per\milli\meter} with $\nu_q=0.1$.
This is closer to the experimentally reported propagation field at \SI{1}{\bar} pressure which is \SIrange{1.2}{1.3}{\kilo\volt\per\milli\meter} \cite{Seeger2017}.

Next, computer simulations using a slightly higher applied voltage of $V=\SI{30}{\kilo\volt}$ showed that streamers failed to develop with $\nu_q=\num{0.01}$, but fully developed using $\nu_q=\num{0.1}$.
\Fref{fig:PositivePhotoi} shows the evolution of this discharge, where the baseline simulation ($\nu_q=\num{1}$) is included for the sake of comparison.
For the simulation with $\nu_q=\num{0.1}$ the positive streamer propagates at about half the velocity of the baseline simulation, corresponding to a velocity of approximately \SIrange{0.1}{0.2}{\milli\meter\per\nano\second}.
The discharge is also highly irregular with a higher plasma density in the filaments, which lies in the range $n_\e= \SIrange[per-mode=reciprocal]{E21}{E22}{\per\cubic\meter}$.
In the baseline simulation the plasma density in the channels was typically \SIrange[per-mode=reciprocal]{E20}{E21}{\per\cubic\meter}.
The corresponding field at the positive streamer tips were also higher, ranging up to \SI{2000}{\td} for the filaments with the smallest diameters.

\begin{figure}[htb]
  \centering
  \includegraphics{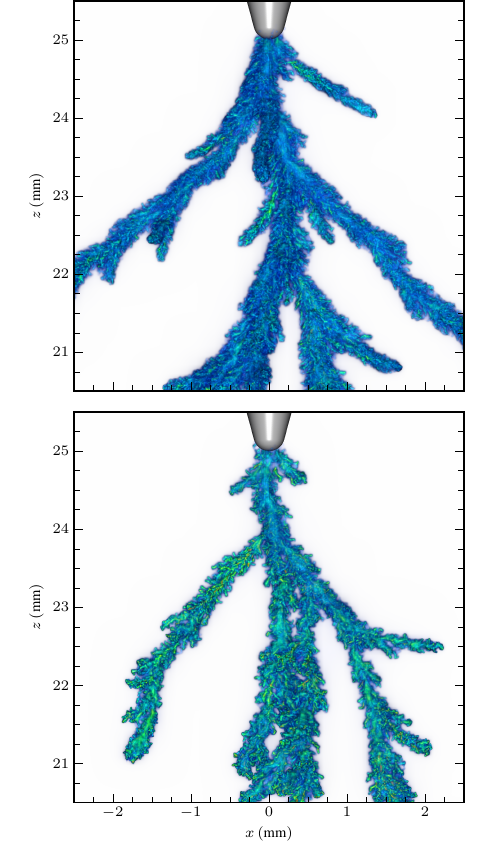}
  \caption{
    Closeup of the irregularity of positive streamers using an applied voltage $V=\SI{30}{\kilo\volt}$ after $t=\SI{40}{\nano\second}$, using $\nu_q=\num{1}$ (top) and $\nu_q=\num{0.1}$ (bottom).
  }
  \label{fig:PositiveZoom}
\end{figure}

Another difference between the streamer evolution with $\nu_q=\num{0.1}$ and $\nu_q=\num{1}$ is that as the amount of photoionization is lowered the streamer grows even more irregularly.
A qualitative demonstration of this is given in \fref{fig:PositiveZoom} which shows a closeup of the plasma density in the vicinity of the anode.
The baseline data ($\nu_q=\num{1}$) in this figure corresponds to the data at $t=\SI{40}{\nano\second}$ in \fref{fig:PositivePhotoi}.
Electron density fluctuations can be seen for the simulation with $\nu_q=\num{1}$, which manifests as a fuzziness in the plasma density in the channels.
This irregularity is due to individually incoming avalanches, caused by stochastic fluctuations due to photoionization at the front of the discharge.
For the simulation with $\nu_q=\num{0.1}$ this feature is much more pronounced, and individual avalanches can be identified.

\paragraph{The role of photoionization in \cotwo}
Despite our initial underestimation of the positive streamer propagation field, we find that positive streamers in \cotwo require substantially higher background fields than positive streamers in air.
The breakdown field in \cotwo is around \SI{86}{\td} while in air the corresponding value is approximately \SI{120}{\td}, so the breakdown field of \cotwo is only \SI{71}{\percent} that of air.
However, positive streamers in atmospheric air can propagate if the average background field is greater than \SI{0.5}{\kilo\volt\per\milli\meter}, but positive streamers in \cotwo require background fields greater than \SI{1}{\kilo\volt\per\milli\meter}.
As we showed, the internal field in positive streamers channels is about the same in \cotwo and air, which implies that positive streamers of similar lengths in air and \cotwo carry the same electric potential at their tips.
The comparatively large difference in propagation fields in these two gases must therefore be due to conditions at the streamer head rather than conditions inside the streamer channels.
Our calculations showed that an increasingly higher voltage is required for propagating positive streamers when the amount of photoionization is lowered, which suggests that the higher propagation field for positive \cotwo streamers could be due to photoionization mechanisms at the streamer tip.
Production of ionizing photons in \cotwo is, relatively speaking, lower than in air.
Furthermore, in \cotwo most of the ionizing photons are absorbed very close to the streamer head, with a mean absorption distance on the order of \SI{30}{\micro\meter}.
Ionizing photons in air propagate longer before they are absorbed, up to \SI{2}{\milli\meter}.
This difference in absorption length implies that free electrons that appear due to ionizing radiation in air multiply exponentially over a longer distance than corresponding free electrons in \cotwo.
Since positive streamers grow due to incoming electron avalanches, and the size of these avalanches depend on where the free electrons initially appeared, a shorter absorption length effectively leads to a weaker photoionization coupling.

Experimentally, \textcite{Seeger2017} found that the pressure-reduced streamer propagation field for negative \cotwo streamers is constant.
In other words, if the pressure is doubled then the minimum applied voltage that is necessary in order to initiate and propagate negative streamers is also doubled.
For positive streamers, however, the authors observed the same behavior that is commonly observed in air: The pressure-reduced streamer propagation field grows with pressure.
This implies that if the pressure is doubled, the applied voltage must be more than doubled in order to initiate and propagate a positive streamer discharge.
Based on the observations made above, we conjecture that this is caused by a reduction in the photoionization level at higher pressures, potentially due to collisional quenching of the EUV-emitting fragments involved in the photoionization process.

\section{Conclusion}
\label{sec:conclusion}

\subsection{Summary}
We have presented a 3D computational study of positive and negative streamer discharges in pure \cotwo, using a microscopic drift-diffusion particle model based on kinetic Monte Carlo.
From the transport data we showed that the existence of a local maximum in the effective attachment rate affects the conductivity, and thus electric field, in the streamer channels.
The reduction in the conductivity leads to a corresponding increase in the electric field.
This occurs on the timescale of \SI{20}{\nano\second} for $E/N\approx \SI{70}{\td}$ at atmospheric pressure, and thus took place on the time scale of our computer simulations.
We suggest that this mechanism is analogous to the attachment instability in air \cite{DouglasHamilton1973}, and that it may play an important role in the further evolution of the discharge.
In air, the attachment instability is associated with increased optical emission and presumably also increased localized heating in the channel, which to a coarse approximation is given by $\bm{E}\cdot\bm{J}$ where only $\bm{J}$ is constant through a channel.
While we have not modeled optical emission nor heating, it is known that the attachment instability is responsible for the long-term optical emission of sprite discharges in the Earth atmosphere (e.g., column glows and beads) \cite{Luque2016, marskar2023genesis}.
Analogous emissions may thus exist for sprites in atmospheres mainly composed of \cotwo, such as the Venusian atmosphere.

In the computer simulations we observed very high electric fields $E/N \sim \SI{2000}{\td}$ in the cathode sheath and on some positive streamer tips (in particular for lower photoionization levels).
We also used a very fine spatial resolution $\Delta x\lesssim \SI{1}{\micro\meter}$.
At these conditions, a standard Courant condition for the maximum permissible time step in fluid-based methods is $\Delta t \leq \Delta x/\left(\left|v_x\right| + \left|v_y\right| + \left|v_z\right|\right)$, which would imply using time steps below $\SI{0.5}{\pico\second}$.
This is $\num{20}$ times shorter than the actual time step we used in our calculations, and would imply taking over \num{100000} time steps in our calculations, which would lead to prohibitively expensive calculations.
However, the particle-based LFA model does not have a Courant condition, and alleviated the need for such a small time step. 
A partial reason for the success of this study was due to this feature, as it allowed us to obtain self-consistent solutions for comparatively large 3D streamer discharges without incurring unacceptable computational costs.

\subsection{Negative streamers}
For negative streamers we obtain a satisfactory agreement with the experiments by \textcite{Seeger2017}.
Our calculations indicate that the negative streamer propagation field at \SI{1}{\bar} pressure is \SIrange{1.1}{1.2}{\kilo\volt\per\milli\meter}.
The streamer velocity is voltage dependent and ranges between \SIrange{0.3}{0.6}{\milli\meter\per\nano\second} in our simulations.
The minimum streamer diameter was at least \SI{420}{\micro\meter}.
Velocities, propagation fields, and radii were in good agreement with experiments.
The channel field in negative \cotwo streamer could become quite high, exceeding \SI{70}{\td} in the channel, which we suggested was due to the attachment instability.
Photoionization was shown to be negligible at the negative streamer head, but nonetheless had a major impact on the streamer evolution since it affects the connection of the negative streamer to the cathode.

\subsection{Positive streamers}
For positive streamers we obtained partial agreement with experiments \cite{Seeger2017}.
The reported streamer propagation field for positive streamers at \SI{1}{\bar} pressure is approximately \SIrange{1.2}{1.3}{\kilo\volt\per\milli\meter}, whereas our baseline calculations gave a propagation field of \SIrange{0.8}{1}{\kilo\volt\per\milli\meter}.
This discrepancy indicates that we probably overestimated the amount of photoionization in our calculations, which could either be due to disregard of collisional quenching, lack of reliable photoionization data, or extrapolation of the available photoionization data outside of the experimentally obtained range.
Although we were unable to answer which of these factors were incorrect, we found that higher voltages were required in order to sustain positive streamer propagation as we reduced the amount of photoionization.

The reported photoionization values by \textcite{Przybylski1962} are sufficient for sustaining positive discharges in \cotwo at \SI{1}{\bar}, despite the fact that photoionization is weaker than in air and that the photons are absorbed very close to the streamer head.
The smallest observed positive streamer radius in the computer simulations was approximately \SI{140}{\micro\meter}, and streamers with this radius did not branch, so they might correspond to the minimal-diameter streamers observed by \textcite{Seeger2017}.
The positive streamers were slower than the negative streamers, with typical velocities in the range \SIrange{0.2}{0.4}{\milli\meter\per\nano\second}.
This is contrary to behavior in air where positive streamers propagate faster than negative streamers \cite{Briels2008a}.
In \cotwo, simulations and experiments \cite{Seeger2017} both show that negative streamers are faster than positive streamers.

\subsection{Outlook}
We discussed the lack of reliable photoionization data for \cotwo, which is in contrast to air where the photoionization process is well identified, and even simplified models provide sufficient accuracy \cite{Wang2023}.
Part of the reason for this situation is the lack of energy-resolved emission cross sections for the fragmented products that appear when \cotwo molecules are dissociated through electron impact.
Experimentally, data is only available at low pressures and it is not known if the reported data by \textcite{Przybylski1962} represents total photoionization, or photoionization per steradian \cite{Pancheshnyi2015}.
In the latter case, the photoionization efficiency $\xi\left(E/N\right)$ that we used in this paper needs to be multiplied by a factor of $4\pi$.
The usage of our parameter $\nu_q$ is then reinterpreted to include the factor of $4\pi$ and the role of quenching. 
Lack of data on the role of collisional quenching of the EUV-emitting fragments was artificially compensated for by reducing photoionization levels by a factor of $10$ and $100$, respectively.
We observed that positive streamers could develop even at such low levels of photoionization, but that their initiation required a higher applied voltage.
We speculate that this is at least partially the reason why experiments show that positive polarity is the dominant breakdown mechanism in \cotwo below atmospheric pressure, while negative polarity dominates at higher pressure \cite{Seeger2017}.

\section*{Acknowledgements}
This study was partially supported by funding from the Research Council of Norway through grant 319930.
The computations were performed on resources provided by UNINETT Sigma2 - the National Infrastructure for High Performance Computing and Data Storage in Norway.

\section*{Data availability statement}
The calculations presented in this paper were performed using the chombo-discharge computer code \cite{Marskar2023} (git hash 540f105).
Although the results included in this paper are stochastic, the input scripts containing simulation parameters and chemistry specifications that were used in this paper are available in both the chombo-discharge results repository at \url{https://github.com/chombo-discharge/discharge-papers/tree/main/ItoKMC-CO2}, and at the following URL/DOI: \url{https://doi.org/10.5281/zenodo.10219863}.

\printbibliography

\end{document}